\documentclass[aps,prc,twocolumn,superscriptaddress]{revtex4-1}

\usepackage[english]{babel}
\usepackage{graphicx}
\usepackage{color}
\usepackage{amssymb}
\usepackage{amsmath}
\usepackage{enumitem}
\usepackage{array}

\renewcommand{\vec}[1]{\mbox{\boldmath$#1$}}
\newcommand{\Elab}{E_\mathrm{lab}}

\begin{document}

\title{Systematic Investigation of Negative Cooper-Frye Contributions in Heavy
  Ion Collisions Using Coarse-grained Molecular Dynamics}

\author{D.~Oliinychenko}
\email{oliiny@fias.uni-frankfurt.de}
\affiliation{Frankfurt Institute for Advanced Studies, D-60438
  Frankfurt am Main, Germany}
\affiliation{Bogolyubov Institute for Theoretical
  Physics, Kiev 03680, Ukraine}

\author{P.~Huovinen}
\email{huovinen@th.physik.uni-frankfurt.de}

\author{H.~Petersen}
\email{petersen@fias.uni-frankfurt.de}
\affiliation{Frankfurt
  Institute for Advanced Studies, D-60438 Frankfurt am Main, Germany}
\affiliation{Institut f\"ur Theoretische Physik,
  Goethe-Universit\"at, D-60438 Frankfurt am Main, Germany}

\begin{abstract}

In most heavy ion collision simulations involving relativistic
hydrodynamics, the Cooper-Frye formula is applied to transform the
hydrodynamical fields to particles. In this article the so-called
negative contributions in the Cooper-Frye formula are studied using a
coarse-grained transport approach. The magnitude of negative
contributions is investigated as a function of hadron mass, collision
energy in the range of $E_{\rm lab} = 5$--$160A$ GeV, collision
centrality and the energy density transition criterion defining the
hypersurface. The microscopic results are compared to negative
contributions expected from hydrodynamical treatment assuming local
thermal equilibrium.

The main conclusion is that the number of actual microscopic particles
flying inward is smaller than the negative contribution one would
expect in an equilibrated scenario.  The largest impact of negative
contributions is found to be on the pion rapidity distribution at
midrapidity in central collisions. For this case negative
contributions in equilibrium constitute 8--13\% of positive
contributions depending on collision energy, but only 0.5--4\% in
cascade calculation. The dependence on the collision energy itself is
found to be non-monotonous with a maximum at 10-20$A$ GeV.

\end{abstract}

\maketitle

\section{Introduction}

Relativistic hydrodynamics is nowadays the standard approach for
modeling ultrarelativistic heavy-ion collisions at highest RHIC
(Relativistic Heavy Ion Collider) and LHC (Large Hadron Collider)
energies. These dynamical descriptions are either based on
ideal~\cite{Kolb:2003dz,Huovinen:2006jp} or dissipative
hydrodynamics~\cite{Heinz:2013th,Gale:2013da} and describe the entire
expansion fluid dynamically. In so called hybrid
approaches~\cite{Hirano:2012kj,Petersen:2014yqa} only the early hot
and dense stage of the expansion is described using hydrodynamics and
the later dilute stage by hadron transport.

Most of these models use a conceptually similar procedure: Given an
initial condition, the hydrodynamic equations are solved in the whole
forward light cone. Near the boundary of vacuum and at the late times
of evolution hydrodynamics is not applicable any more, when the
density is small and the mean free path is larger than the system
size. Therefore, models switch to an off-equilibrium microscopic
description in terms of particles in this region. In hybrid approaches
particles can scatter, while other models allow only free-streaming
and resonance decays. In any case, the most commonly used way to
convert the fluid-dynamical fields to particles, a process that we
call here 'particlization', is by using the Cooper-Frye formula.

The Cooper-Frye formula assumes particlization to take place on
infinitesimally thin three-dimensional hypersurface in four-dimensional
space-time. This hypersurface $\Sigma$ is usually determined a
posteriori from hydrodynamical solution in the whole forward light
cone, usually as a hypersurface of constant time, energy density,
temperature, or Knudsen number. Particle distributions on an
infinitesimal element of hypersurface, $d\Sigma$, are calculated using the
following formula:
\begin{eqnarray} \label{CF_formula}
p^0 \frac{d^3N}{d^3p} = p^{\mu} d\sigma_{\mu} f(p) \,,
\end{eqnarray}
where $f(p)$ is a distribution function and $d\sigma_{\mu}$ a normal
four-vector of hypersurface with length equal to the area of the
infinitesimal surface element. This formula was obtained by Cooper and
Frye~\cite{ref:CF74} with the main feature that it respects
four-momentum conservation. Though formula (\ref{CF_formula}) is valid
for any $f(p)$, the distribution function is usually assumed to be
either the boosted thermal distribution $f(p) = f_0(p) = \left[ exp
  \left(\frac{p^{\mu}u_{\mu} - \mu}{T} \right) \pm 1 \right]^{-1}$
(ideal fluid), or a distribution close to the boosted thermal
distribution $f(p) = f_0(p) + \delta f(p)$ (viscous fluid), where
$\delta f(p)$ is the dissipative correction. Here $T$, $\mu$ and
$u^\mu=\gamma(1,\vec{v})$ are temperature, chemical potential and the
flow velocity of the fluid, respectively.

There is, however, a conceptual problem with the Cooper-Frye
formula. Where the surface is space-like, \emph{i.e.}, its normal
vector $d\sigma_{\mu}$ is space-like, and $p^\mu d\sigma_\mu < 0$ for
some $\vec{p}$. Thus if $f(p)>0$ for all $p$, as is the case for the
thermal distribution, $\frac{d^3N}{d^3p} < 0$ for some $\vec{p}$. This
can be easily seen in the local rest frame of a space-like surface
(which always exists since $v_{surf} < c$ for space-like surfaces),
where $p^{\mu} d\sigma_{\mu} = \vec{p} \cdot \vec{n}$ and thus
$\frac{d^3N}{d^3p} < 0$ for momenta directed inward the surface. On
the other hand, for those time-like surfaces which normal vector
points toward the future (\emph{i.e.}, $d\sigma_0 > 0$),
$\frac{d^3N}{d^3p} > 0$ for any $\vec{p}$. This can be also understood
as follows: surface is ''escaping'' faster than the speed of light, so
no particle can cross it inward. (For a summary of the properties of
time-like and space-like surfaces, see Table~\ref{Tab:surf_prop}).

\begin{table}[ht]
 \begin{tabular}{cc}
   \hline \hline
   time-like surface & space-like surface  \\
   \hline
   time-like normal & space-like normal \\
   $d\sigma^{\mu}d\sigma_{\mu}>0$ & $d\sigma^{\mu}d\sigma_{\mu}<0$ \\
   $v_{surf} > c$  &  $v_{surf} < c$ \\
   $\exists$ RF: $d\sigma^{\mu} = (\pm dx\,dy\,dz, 0, 0, 0)$ &
   $\exists$ RF: $d\sigma^{\mu} = (0, 0, 0, dt\,dx\,dy)$ \\
   $d\sigma_0 > 0 \Rightarrow \forall p^{\mu}$: $p^{\mu}d\sigma_{\mu} > 0$ &
   $\exists p^{\mu}$: $p^{\mu}d\sigma_{\mu} < 0$ \\
   $d\sigma_0 > 0 \Rightarrow \forall p^{\mu}$: $\frac{d^3N_{CF}}{d^3p} > 0$ &
   $\exists p^{\mu}$: $\frac{d^3N_{CF}}{d^3p} < 0$ \\
   \hline \hline
 \end{tabular}
 \caption{Properties of surface elements. $g^{\mu \nu} =
   (1,-1,-1,-1)$. The normal vector is directed toward lower
   density. RF abbreviates Reference Frame, $\frac{d^3N_{CF}}{d^3p}$
   denotes particle distribution from the hypersurface element calculated
   using the Cooper-Frye formula. }
 \label{Tab:surf_prop}
\end{table}

If $\frac{d^3N}{d^3p}$ is interpreted as a phase-space density,
negative values of it are clearly unphysical, but instead of giving a
literal phase-space density, Cooper-Frye formula rather counts the
world lines of particles crossing the surface element $d\Sigma$, and
gives positive weight to particles moving ``outward'' and negative
weight to particles moving ``inward''. Thus the negative values of
$\frac{d^3N}{d^3p}$, the so-called negative Cooper-Frye contributions,
refer to particles flying inward toward the hydrodynamical region, and
which should thus be absorbed back to the fluid. 

In pure hydrodynamical models, this poses a problem: Particlization
takes place at freeze-out when rescatterings cease, and particles
stream free. Thus, once particles cross the particlization surface,
there is nothing from where particles could scatter back toward the
surface, and thus there should be no particles flying back. To avoid
this problem, one could choose a completely time-like particlization
hypersurface, for example a hypersurface of a constant time without
any negative contributions. However, it was shown~\cite{ref:Rish98}
that particle spectra obtained in such an approach are dramatically
different from spectra on a constant temperature hypersurface. Another
way is to consider cut-off distribution~\cite{ref:Bugaev96}: 
$p^0 \frac{d^3N}{dp^3} = 
 p^{\mu} d\sigma_{\mu} f(p) \Theta (p^{\mu} d\sigma_{\mu})$. Such a
prescription violates conservation laws, unless one adjusts
temperature, chemical potentials, and flow velocity in the particle
distribution $f(p)$~\cite{Anderlik:1998cb,Anderlik:1998et}.

On the other hand, there is no such a problem in hybrid models.
Particlization takes place where rescatterings are abundant, and thus
it is natural to have particles flying back to the fluid-dynamical
region. The problem is rather a practical one: What does the negative
weight of a particle mean when one samples the particle distributions
at particlization surface to create an initial state for the hadron
transport? Usually one simply ignores them (see \emph{e.g.}
Ref.~\cite{ref:PetHuo12}), which violates conservation laws. An
attempt to include these negative weights to the hadron transport was
recently made in Ref~\cite{ref:Pratt14}. Alternatively, if the
transition from fluid to transport takes place in a region where
hydrodynamics and transport are equivalent, the negative Cooper-Frye
contributions coincide with the distribution of particles that
backscatter to hydrodynamical region. Thus all one needs to do is to
remove these particles from the cascade, but such removing is
technically challenging, and the problem remains how to find the
region where hydrodynamics and transport lead to equal
solutions---assuming that such a region exists at all! Thus the
ultimate solution to the problem would be to construct a model,
solving coupled hydrodynamical and kinetic equations with the kinetic
model providing boundary condition for hydrodynamics. An attempt in
this direction was taken by
Bugaev~\cite{ref:Bugaev99,ref:Bugaev02,ref:Bugaev04}, but these ideas
have not yet been implemented in practice.

Fortunately, at high collision energies, the explosive expansion
dynamics keeps the negative contributions on the level of a few
percent. Emission of particles from time-like areas of surface where
no negative contributions appear (so-called volume emission) is much
larger than emission from space-like areas (so-called surface
emission), and as we will discuss later, large flow velocity reduces
negative contributions from space-like surfaces. Nevertheless, there
are very few studies that actually quote the values of negative
contributions, and investigations at lower collision energies are
lacking completely. In this article the negative contributions arising
on the Cooper-Frye transition surface assuming distribution functions
in local equilibrium are compared to the actual underlying microscopic
dynamics to investigate the systematic differences between a transport
and a hybrid approach.

Therefore, the aim of the current study is to compare the expected
negative contributions in a locally equilibrated hydrodynamical
approach with the actual number of particles that scatter back through
a hypersurface in a coarse-grained microscopic transport approach. A
constant energy density transition surface is constructed and negative
Cooper-Frye contributions are compared to actual backscattered
particles. In addition, the magnitude of negative contributions is
calculated in a systematic way depending on hadron sort, collision
energy, centrality, and choice of the transition surface. In Section
\ref{sec:Methodology} the framework for the calculation is
explained. Section \ref{sec:Tests} shows results of tests of the
numerical setup and sensitivity to internal parameters of the
calculation. Finally, Section \ref{sec:Results} contains physical
results: the quantification of Cooper-Frye negative contributions and
their comparison to backscattered particles.

\section{Methodology}
\label{sec:Methodology}

Our calculation is based on the hadronic transport approach -
Ultrarelativistic Quantum Molecular Dynamics (UrQMD 3.3p2)
\cite{ref:UrQMD}. The degrees of freedom in UrQMD are hadrons,
resonances up to a mass of 2.2 GeV and strings and the implemented
processes include binary elastic and inelastic scatterings which
mainly proceed via resonance formation and decays or string excitation
and fragmentation at higher collision energies. The UrQMD particles
move along classical trajectories and scatter according to their
free-particle cross-sections. In our studies there are no long range
potentials and particle trajectories between collisions are always
straight lines. Using UrQMD we simulate Au + Au collisions at
laboratory frame energies $E_{\rm lab} =$ 5, 10, 20, 40, 80 and 160$A$
GeV. This energy region is chosen because we expect UrQMD to provide a
reasonable description of the collision dynamics at those energies,
and the Cooper-Frye negative contributions to become significant in
this energy range.

The general procedure for our calculations is:
\begin{enumerate}
\item Generate many UrQMD events and coarse-grain them using a 3+1D
  space-time grid.
\item Find the local energy density in the Landau rest frame of each
  grid cell, $\epsilon_{LRF}(t,x,y,z)$, and the collective flow
  velocity in each cell, $\vec{v}(t,x,y,z)$.
\item Construct the hypersurface $\Sigma$ of a constant energy density
  $\epsilon_{LRF}(t,x,y,z) = \epsilon_c$.
\item Calculate the particle spectra on $\Sigma$ by using the
  Cooper-Frye formula and by counting the actual UrQMD particles that
  cross $\Sigma$.  To obtain these spectra and to compare them to each
  other is the goal of the current work.
\end{enumerate}
This procedure mimics switching from hydrodynamics to transport in a
hybrid model, but here the ''hydrodynamical'' picture is obtained by
averaging over particle distributions on a space-time grid. Since all
the information is still available in the underlying microscopic
approach we are able to compare the negative Cooper-Frye contributions
to the spectrum of actual backscattered particles. In the following
we explain all necessary details for each of these steps of the
calculation.

\subsection{Calculating physical quantities on a grid}

To obtain the energy density in the Landau rest frame as a function of
space-time, that is necessary to construct the Cooper-Frye transition
surface, the energy momentum tensor and the net baryon current in the
computational frame are calculated
\begin{eqnarray}
T^{\mu \nu} (t,x,y,z) =\frac{1}{\Delta x \Delta y \Delta z}\left\langle \sum \frac{p^{\mu}p^{\nu}}{p^0} \right\rangle_{\!N} \\
j^{\mu}_B (t,x,y,z) =\frac{1}{\Delta x \Delta y \Delta z}\left\langle \sum \frac{p^{\mu} }{p^0} B \right\rangle_{\!N} \,,
\end{eqnarray}
where the sum is over all particles in each grid cell at the moment
$t$, and $B$ is the baryon number of each particle. Angular brackets
denote averages over $N$ UrQMD events. The cell sizes need to be small
enough so that gradients of all relevant physical quantities within
the cell are small. On the other hand, if the cell sizes are too small
one needs to generate infeasibly many events to damp statistical
fluctuations of $T^{\mu \nu}$ components from cell to cell, and obtain
a smooth surface $\Sigma$. To satisfy these conditions and to ensure
energy conservation precisely we choose $\Delta x = \Delta y = $ 1 fm,
$\Delta z$ = 0.3 fm and time step $\Delta t$ = 0.1 fm. For the highest
collision energy, $E_{\rm lab} = 160A$ GeV, the gradients are larger,
so even smaller grid sizes were taken: $\Delta x = \Delta y = $ 0.3 fm
and $\Delta z$ = 0.1 fm. This choice is further discussed in the
Section \ref{sec:Tests}, where the sensitivity of results to the grid
size is studied. Since even $N = 10\,000$ events do not provide enough
statistics to obtain a smooth hypersurface, and increase of $N$ is not
feasible due to limited storage capacities, the individual particles
are smeared by marker particles distributed according to a Gaussian
distribution.

Every UrQMD particle with coordinates $(t_p,x_p,y_p,z_p)$ and
four-momentum $p^{\mu}$ is substituted by $N_{split}$ particles with
coordinates distributed with probability density
$f(x,y,z) \sim \exp \left(-\frac{(x-x_p)^2}{2\sigma^2} 
                  - \frac{(y-y_p)^2}{2\sigma^2}
                  - \gamma_z^2 \frac{(z-z_p)^2}{2\sigma^2}\right)$, 
where $\gamma_z = (1-p^z/p^0)^{-1/2}$. These marker particles are
attributed the 4-momentum and quantum numbers of the original particle
divided by $N_{split}$. In our calculation $N_{split}$ = 300 and
$\sigma$ = 1 fm. The sensitivity of our results to the width of the
Gaussian is discussed in Section \ref{sec:Tests}. When this Gaussian
smearing is applied, stable results are obtained with only $N$ = 1500
events, which we employ for our calculations.

\subsection{The hypersurface construction}

After obtaining $T^{\mu \nu}$ in the computational frame, it has to be
transformed to the Landau rest frame (LRF) in each cell. By
definition, $T^{0i}_{LRF} = 0$, \emph{i.e.}, the energy flow in the
LRF is zero.  To find the LRF we solve the generalized eigenvalue
problem $(T^{\mu \nu} - \lambda g^{\mu \nu})h_{\nu} = 0$. The
eigenvector corresponding to the largest eigenvalue is proportional to
the 4-velocity of the LRF and the proportionality constant is fixed by
the constraint that $\sqrt{u_\mu u^\mu}=1$. After finding
$T^{\mu\nu}_{LRF}$ the hypersurface of constant Landau rest frame
energy density is constructed where
$T^{00}_{LRF} \equiv \epsilon_{LRF}(t,x,y,z) = \epsilon_c$, with
$\epsilon_c$ a parameter that characterizes the hypersurface. In such
a way we mimic the transition surface in hybrid models, which
typically use $\epsilon_c = 0.3$--1 GeV/fm$^3$~\cite{ref:PetHuo12}.
The isosurface is constructed using the Cornelius
subroutine~\cite{ref:PetHuo12}, that provides a continuous surface
without holes and avoids double counting of hypersurface pieces. The
subroutine provides the normal four-vectors $d\sigma_{\mu}$ of the
hypersurface. The physical quantities on the grid, \emph{i.e.}, the energy,
net baryon density and the flow velocity, are linearly interpolated to
the geometrical centers of the hypersurface elements.

\subsection{Thermodynamic quantities}

To apply the Cooper-Frye formula one needs the temperature $T$ and
chemical potentials on the surface, which do not exist in the
microscopic picture. Strictly speaking they make sense only in the
vicinity of thermal and chemical equilibrium, which may not be the
case in our UrQMD simulation. Nevertheless, we take the LRF energy
density and net baryon density to mean equilibrium densities---as is
the case when deviations from equilibrium are small---and obtain
temperature and chemical potentials from an ideal hadron resonance gas
(HRG) equation of state (EoS) containing the same hadrons and
resonances as UrQMD. Since our EoS assumes zero strangeness density,
we impose this constraint as well, even if UrQMD itself allows local
non-zero strangeness. In practice, this means solving the following
coupled equations to find the temperature $T$, baryon chemical
potential $\mu_B$ and strangeness chemical potential $\mu_S$:
\begin{eqnarray}
\label{Eq:EoS1}
\epsilon_{LRF} &=& \sum_p \frac{g_p}{(2\pi)^3} \int 
             \frac{d^3k \, \sqrt{k^2 +m^2} }{e^{(\sqrt{k^2 + m_p^2}-\mu_B B_p  - \mu_S S_p)/T} \pm 1} \\
\label{Eq:EoS2}             
n_B^{LRF} &=&  \sum_p \frac{g_p B_p}{(2\pi)^3} \int 
             \frac{d^3k }{e^{(\sqrt{k^2 + m_p^2}-\mu_B B_p  - \mu_S S_p)/T} \pm 1} \\
\label{Eq:EoS3}             
n_S^{LRF} &=&  \sum_p \frac{g_p S_p}{(2\pi)^3} \int 
             \frac{d^3k }{e^{(\sqrt{k^2 + m_p^2}- \mu_B B_p  - \mu_S S_p)/T} \pm 1}             
\end{eqnarray}
Here $\epsilon_{LRF} = T^{00}$ is the energy density in LRF,
$n_B^{LRF}$ is the baryon density in LRF, $n_s$ is the strangeness
density, and the sum runs over all hadron species that appear in
UrQMD; $m_p$ is the mass of a hadron $p$, $g_p$ is its spin and
isospin degeneracy factor, and $B_p$ and $S_p$ are its baryon number
and strangeness, respectively.

\subsection{Cooper-Frye and ''by particles'' calculations} 

After the hypersurface of constant LRF energy density $\Sigma$ is
obtained and $T$ and $\mu$ are evaluated using the EoS, the
Cooper-Frye formula is applied on the hypersurface. The spectrum from
the Cooper-Frye formula is split into positive and negative parts:
\begin{eqnarray} 
\label{Eq:CFdef1}
\frac{dN^{+}_{CF}}{p_T dp_T d\varphi dy } 
  & = & \frac{g}{(2\pi)^3}
         \int_{\sigma}\frac{\Theta(p^{\mu}d\sigma_{\mu}) \, p^{\mu}d\sigma_{\mu}}
         {e^{(p^{\nu}u_{\nu}-\mu)/T} \pm 1} \\
\label{Eq:CFdef2}
\frac{dN^{-}_{CF}}{p_T dp_T d\varphi dy } 
  & = & \frac{-g}{(2\pi)^3}
         \int_{\sigma}\frac{\Theta(-p^{\mu}d\sigma_{\mu}) \,p^{\mu}d\sigma_{\mu} }           
         {e^{(p^{\nu}u_{\nu}-\mu)/T} \pm 1}
\end{eqnarray}
To evaluate $dN/dy$ or $dN/p_T dp_T$ the integrations are performed
numerically, applying the 36$\times$36 points Gauss-Legendre method to
integrals transformed to finite limits.

For comparison with the Cooper-Frye calculation we count the actual
microscopic (not marker) particles crossing the same hypersurface
$\Sigma$ that is used for Cooper-Frye calculations. Inward and outward
crossings are counted separately. To find the point, where a particle
trajectory crosses $\Sigma$ we use the fact that by construction the
energy density $\epsilon > \epsilon_c$ inside the surface and
$\epsilon < \epsilon_c$ outside of it. The energy density is
interpolated to the particle trajectory to find the point where
$\epsilon - \epsilon_c$ changes sign. Each of these crossings is
counted as positive, if the particle streams outward and negative, if
the particle flies toward higher energy densities.

Both Cooper-Frye calculation and particle counting start at the same
time $t_{start}$, which depends on the collision energy. Following the
prescription from hybrid models, we take $t_{start} =
\frac{2R}{v\gamma}$.  This is the time two nuclei need to pass through
each other. Numerical values are 8 fm/c for $5A$ GeV, 5.6 fm/c for
$10A$ GeV, 4 fm/c for $20A$ GeV, 2.8 fm/c for $40A$ GeV, 2 fm/c for
$80A$ GeV and 1.4 fm/c for $160A$ GeV. The same $t_{start}$ is used
for all centralities.

\section{Sensitivity to internal parameters and fulfillment of
  conservation laws}
\label{sec:Tests}

\begin{figure}[htp]
\includegraphics[height=4.6cm]{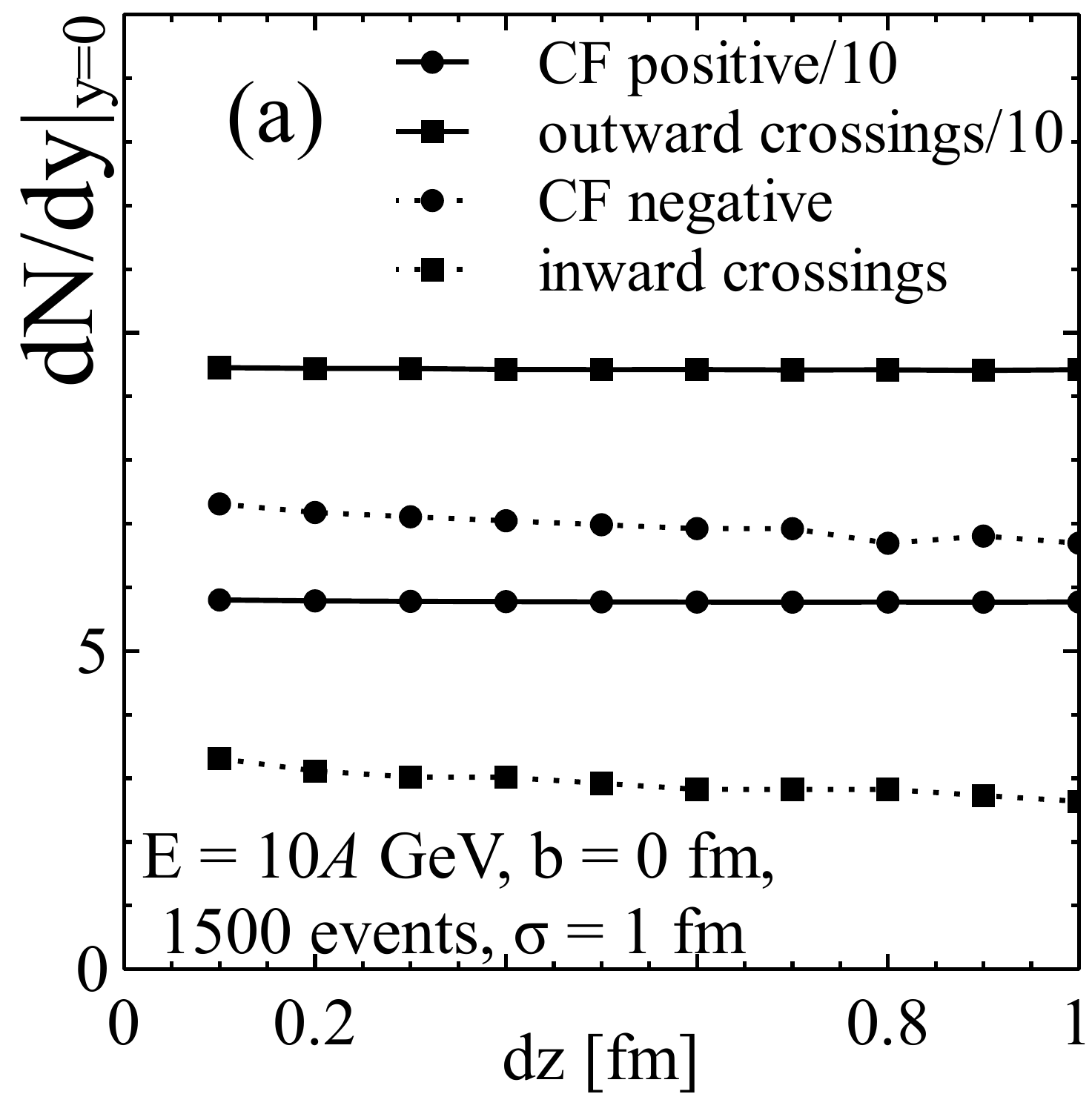}
\includegraphics[height=4.6cm]{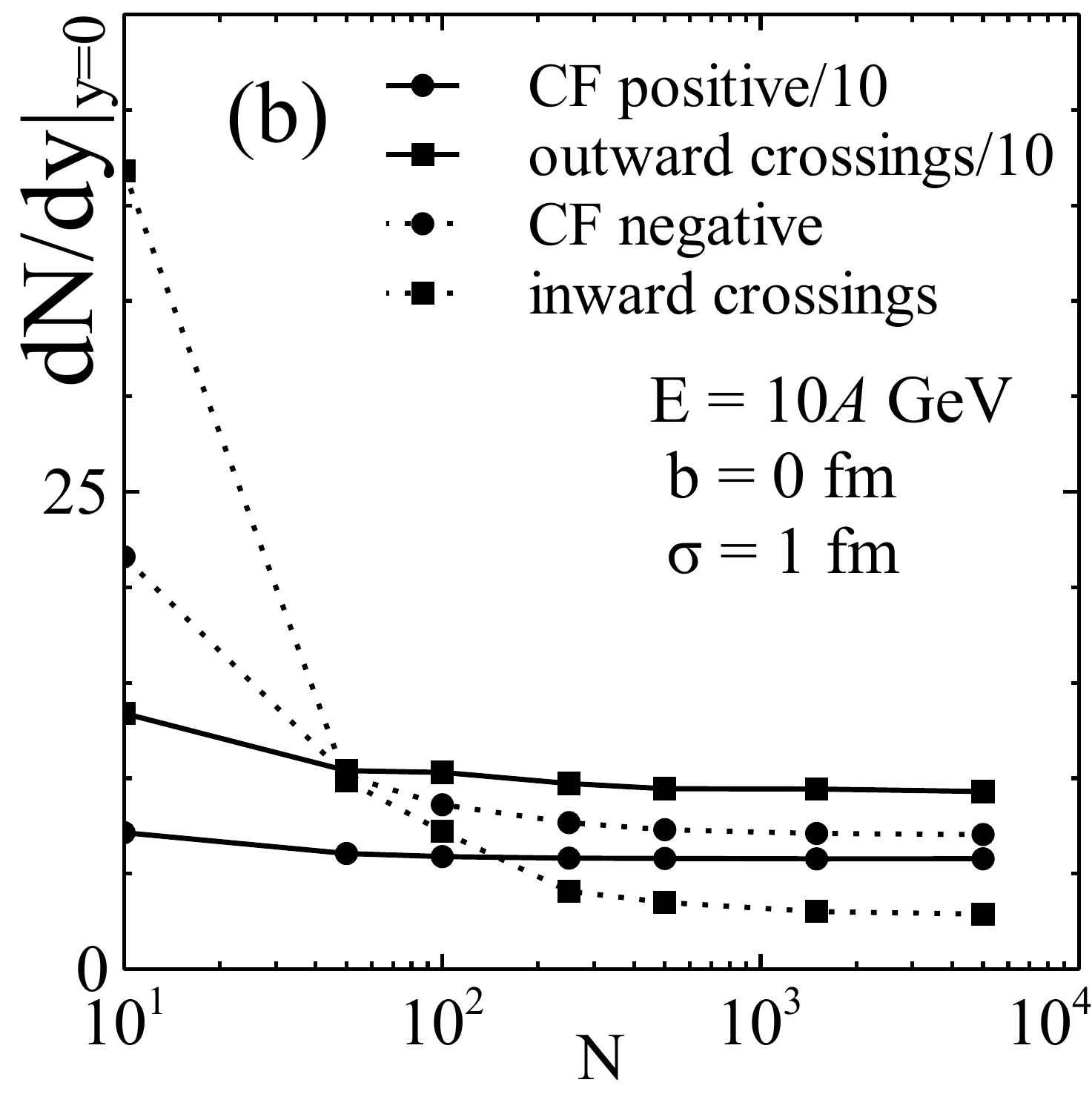}
\includegraphics[height=4.6cm]{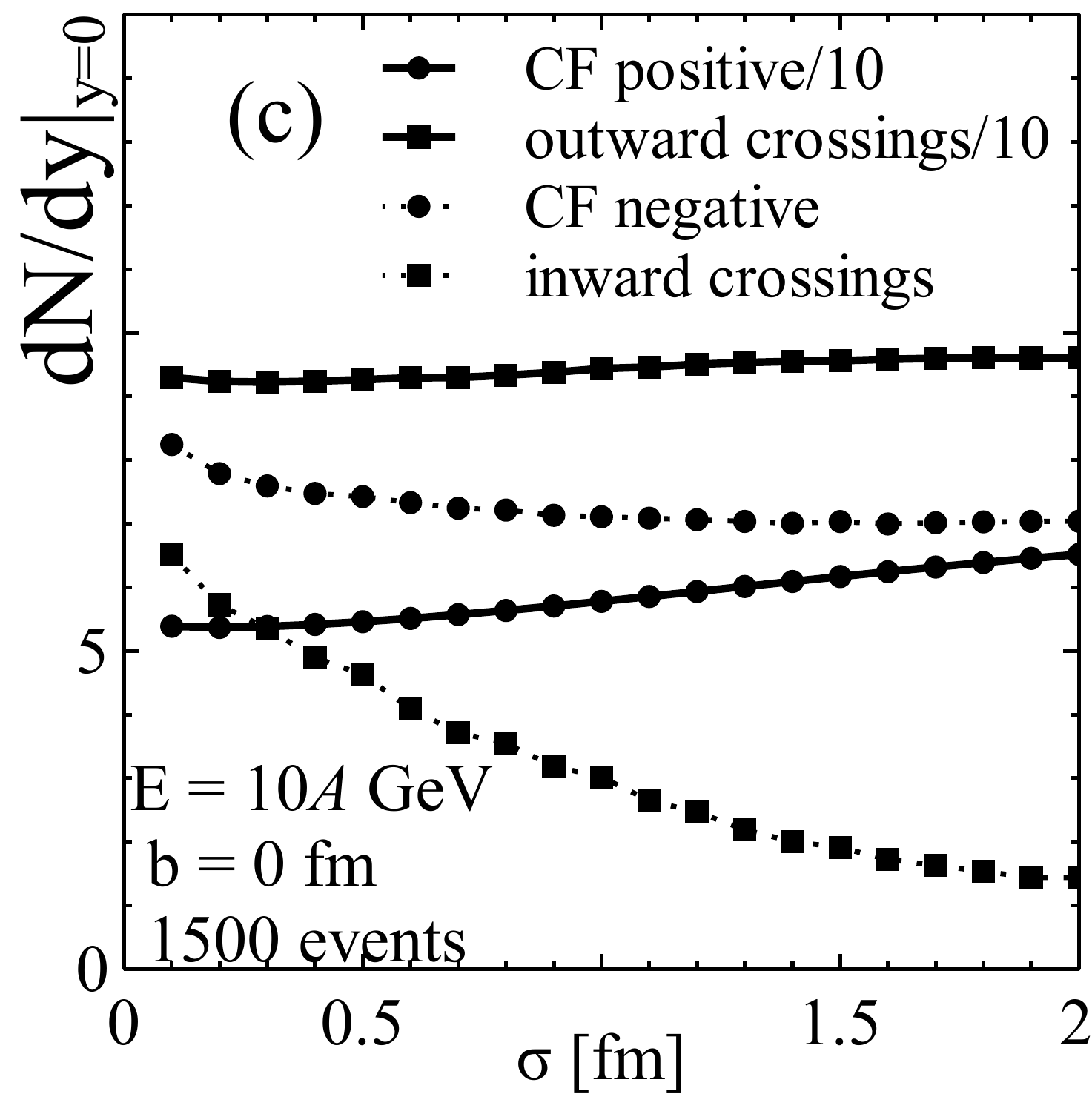}
\caption{Sensitivity of results to internal parameters of the
  simulation: grid spacing along z axis, $dz$ (a), number of
  events, N (b) and the width $\sigma$ of Gaussian
  smearing (c). }
\label{Fig:Int_par}
\end{figure}

\begin{figure*}[htp]
\includegraphics[height=7cm]{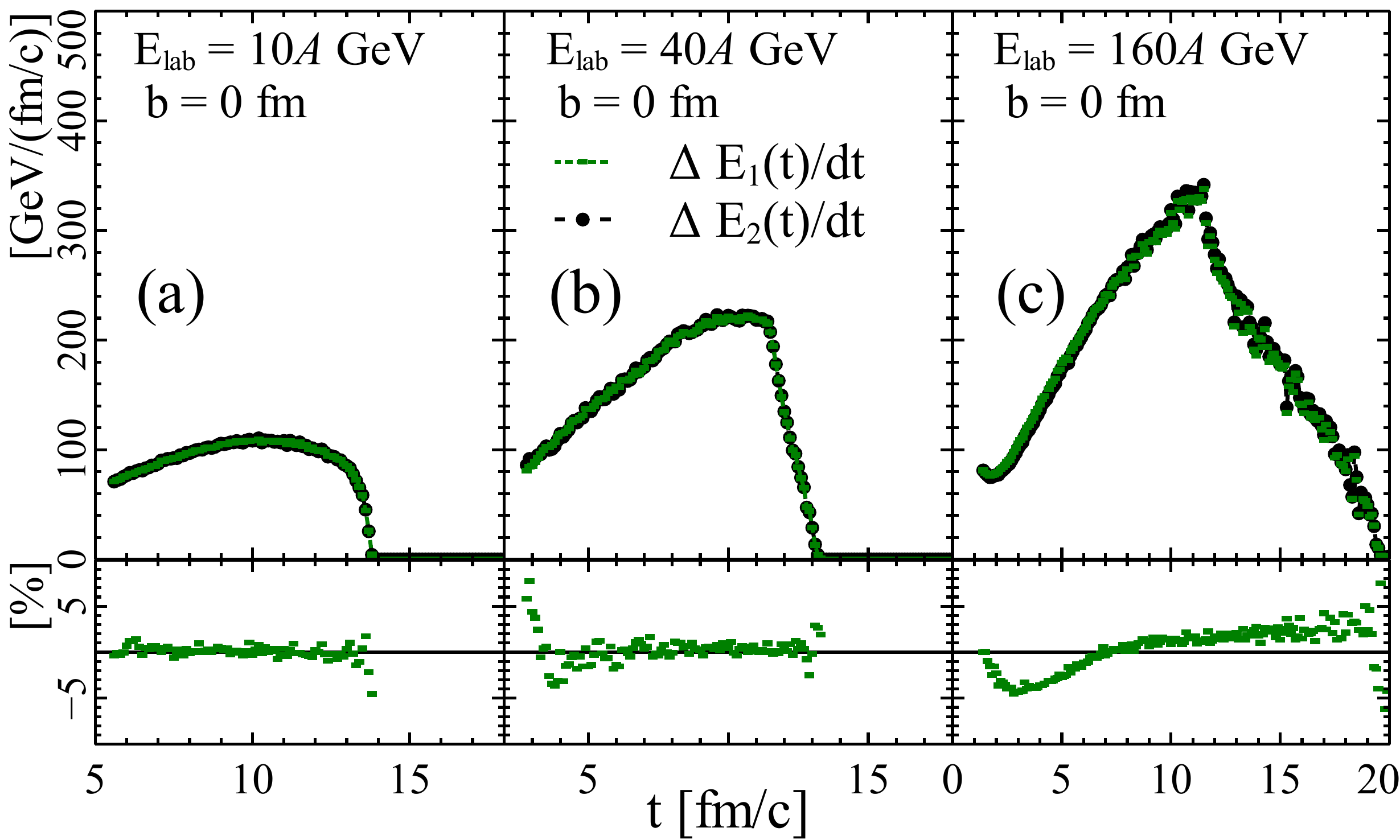}
\caption{(Color online) Energy flux through the surface at different
  times evaluated as actual flow, 
  $\Delta E_1(t)/dt = \int_{t-dt}^{t} T^{\mu 0} d\sigma_{\mu}/dt$
  (circles), and as a difference in energy within the surface at
  different times, $\Delta E_2(t)/dt = (E_{in}(t) - E_{in}(t-dt))/dt$
  (rectangles). Lower panel shows the relative difference between
  these two measures in \%, and thus the conservation of energy in the
  calculation.}
\label{Fig:Econs}
\end{figure*}

Besides physical parameters like the beam energy, $E_{\rm lab}$, and
centrality of the collision controlled by the impact parameter $b$,
our simulation contains internal parameters like grid spacing, the
width of the smearing Gaussian $\sigma$, and the number of events $N$.
Ideally, we should work in such a region of internal parameters, that
our results are independent of them. To see how sensitive our results
are to these internal parameters, the positive and negative
contributions to the pion yield at midrapidity,
$\frac{dN}{dy}|_{y=0}$, at different values of these parameters are
evaluated.

The calculation is more sensitive to the grid spacing in z direction,
$dz$, than to the spacings in x and y directions, $dx$ and $dy$, since
gradients of $T^{\mu\nu}$ are largest in the longitudinal
direction. Although, as shown in Fig.~\ref{Fig:Int_par}~a), even the
sensitivity to $dz$ is weak over a reasonable range of values. The
main motivation for choosing the grid spacing and time step comes in
fact from the requirement of energy conservation discussed later.

The results are very sensitive to the small number of events (see
Fig.~\ref{Fig:Int_par}~b), but already $N = 500$ events provides
sufficient statistics for stable results.  To be on the safe side, we
have analyzed $N=1500$ events for our final results. Unfortunately,
our results are not completely independent of the width $\sigma$ of
the Gaussian smearing, as shown on Fig.~\ref{Fig:Int_par}~c). The
number of inward crossing UrQMD pions is most sensitive to
$\sigma$. Two effects play a role here: at small $\sigma$ the surface
still has large statistical fluctuations and small scale structures,
``lumps'' (See Fig.~2 of Ref~\cite{Huovinen:2002im}), whereas at large
$\sigma$ Gaussian smearing pushes transition surface further out in
space. Further out the densities are smaller, and the UrQMD particle
distributions are further away from equilibrium so that especially the
number of particles moving toward the center is strongly reduced.  We
choose $\sigma = 1$ fm as a reasonable value for our calculations, but
keep in mind that varying $\sigma$ in the range from 0.6 fm to 1.4 fm
causes $\sim$ 20 \% difference in the number of inward crossings. We
consider this a systematic error in our analysis, but fortunately this
uncertainty does not affect our main conclusions.

To check that energy is conserved in the coarse-graining procedure, we
evaluate the energy flow through the surface during the time step
$dt$, $\Delta E_1(t) = \int_{t-dt}^{t} T^{\mu 0} d\sigma_{\mu}$, and
compare it to the change in energy within the surface during the same
time step, $\Delta E_2(t) = E_{in}(t) - E_{in}(t-dt)$, where $E_{in}$
is total energy of particles inside the surface. Ideally 
$\Delta E_1(t) = \Delta E_2(t)$ for any $dt$, but finite cell sizes
limit the precision and break the conservation of energy. The accuracy
of $\Delta E_1 \approx \Delta E_2$ improves when grid spacing and time
step are decreased. Fig.~\ref{Fig:Econs} shows the energy flux
through the surface and the relative difference between
$\Delta E_1(t)$ and $\Delta E_2(t)$ in central collisions at energies
$E_{\rm lab} = 10$, 40, 160$A$ GeV. To achieve better than 5\% percent
accuracy at all times, we use small grid spacing with 
$\Delta x = \Delta y = 1$ fm, $\Delta z$ = 0.3 fm, and time step
$\Delta t$ = 0.1 fm/c in collisions with $E_{\rm lab} \le 80A$ GeV,
and even finer grid with $\Delta x = \Delta y = 0.3$ fm, and 
$\Delta z = 0.1$ fm for collisions at $E_{\rm lab} = 160A$ GeV. When
integrated over the whole collision time, the violation of energy
conservation is less than 1\% at all collision energies. We have done
a similar check for the net baryon charge, and obtained similar
results.

\section{Results and discussion} \label{sec:Results}

\begin{figure}[htp]
\includegraphics[height=6cm]{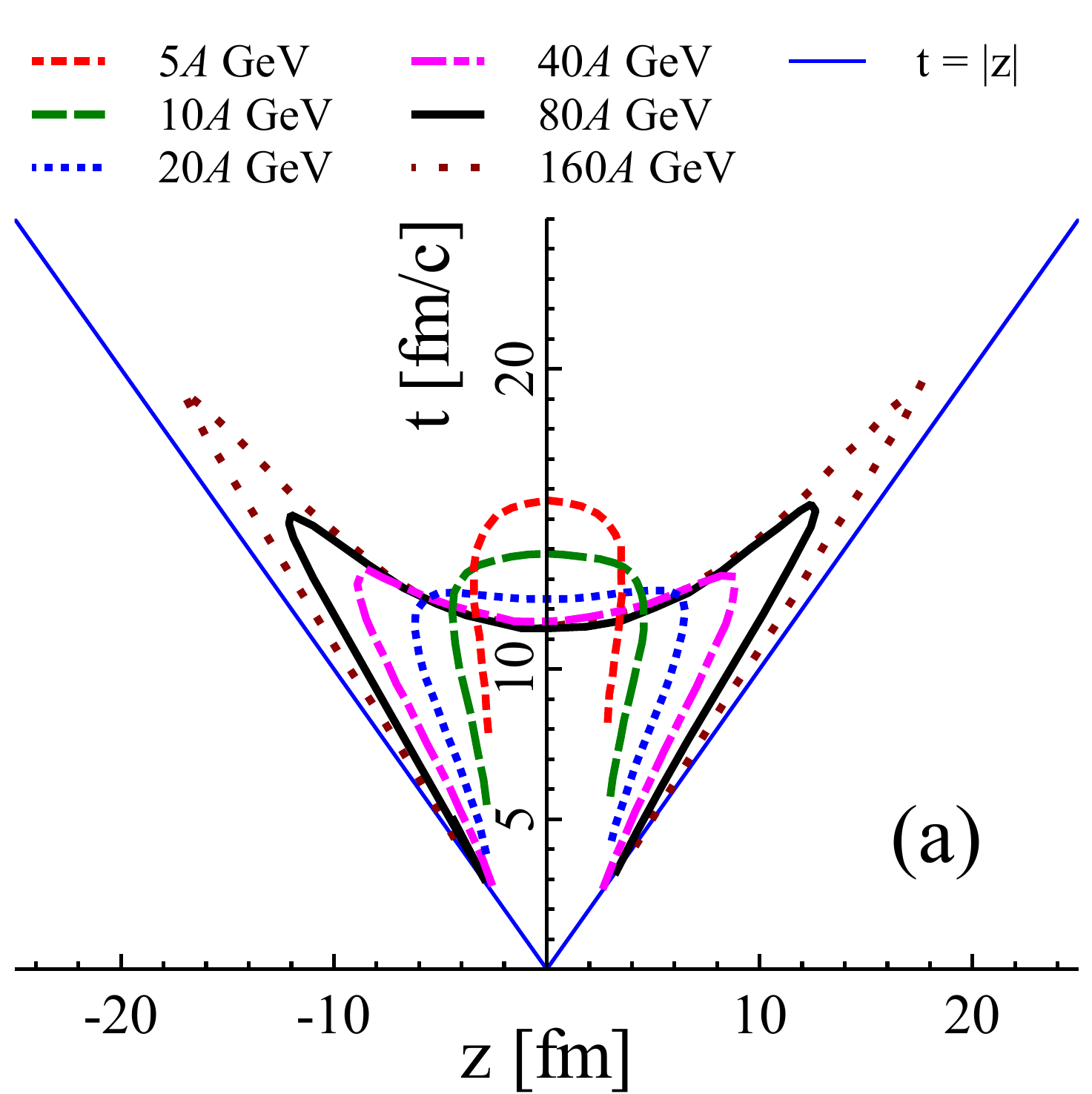}
\includegraphics[height=6cm]{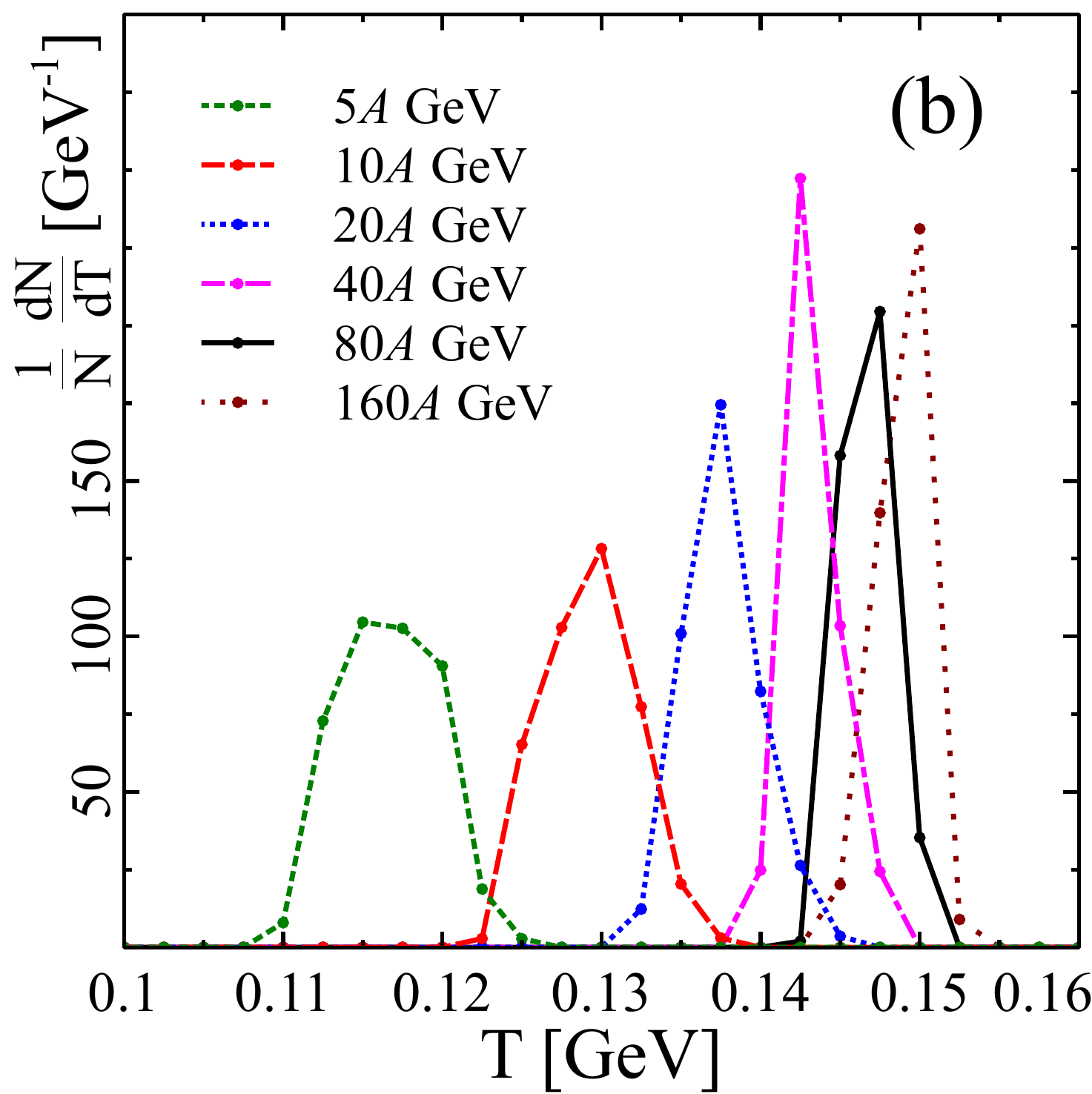}
\caption{(Color online) Upper panel: Hypersurface of constant LRF
  energy density $\epsilon(t,0,0,z) = \epsilon_c = 0.3$
  GeV/fm$^3$. Lower panel: The fraction of hypersurface elements with
  (apparent) temperature $T$ in central Au+Au collisions at the
  collision energy of $E_{\rm lab} = 5$, 10, 20, 40, 80, 160$A$ GeV.}
\label{Fig:surf}
\end{figure}

Let us start by investigating the properties of the transition
hypersurface itself as a function of beam energy. Figure
\ref{Fig:surf} depicts the surface $\Sigma$ in longitudinal direction
along the x axis. We see that with increasing energy, the lifetime of
the system increases. This indicates longer lasting surface emission
(from space-like parts of the surface), which might lead to larger
negative contributions. On the other hand, with increasing energy the
longitudinal expansion leads to larger volume of the final volume
emission (from time-like parts of the surface), which indicates
smaller negative contributions. Thus we have two competing effects,
and one has to carry out the actual calculation to find out how the
negative contributions depend on energy.

Distributions of the (apparent) temperature of the hypersurface
elements are shown on the right panel of Fig.~\ref{Fig:surf}. At each
collision energy temperature distribution is rather narrow, which
means that constant energy density surface approximately coincides
with constant temperature surface. As well, the average temperature
increases with increasing collision energy as expected from thermal
model fits to particle yields~\cite{Andronic:2008gu}.

\begin{figure}[htp]
\includegraphics[width=5cm]{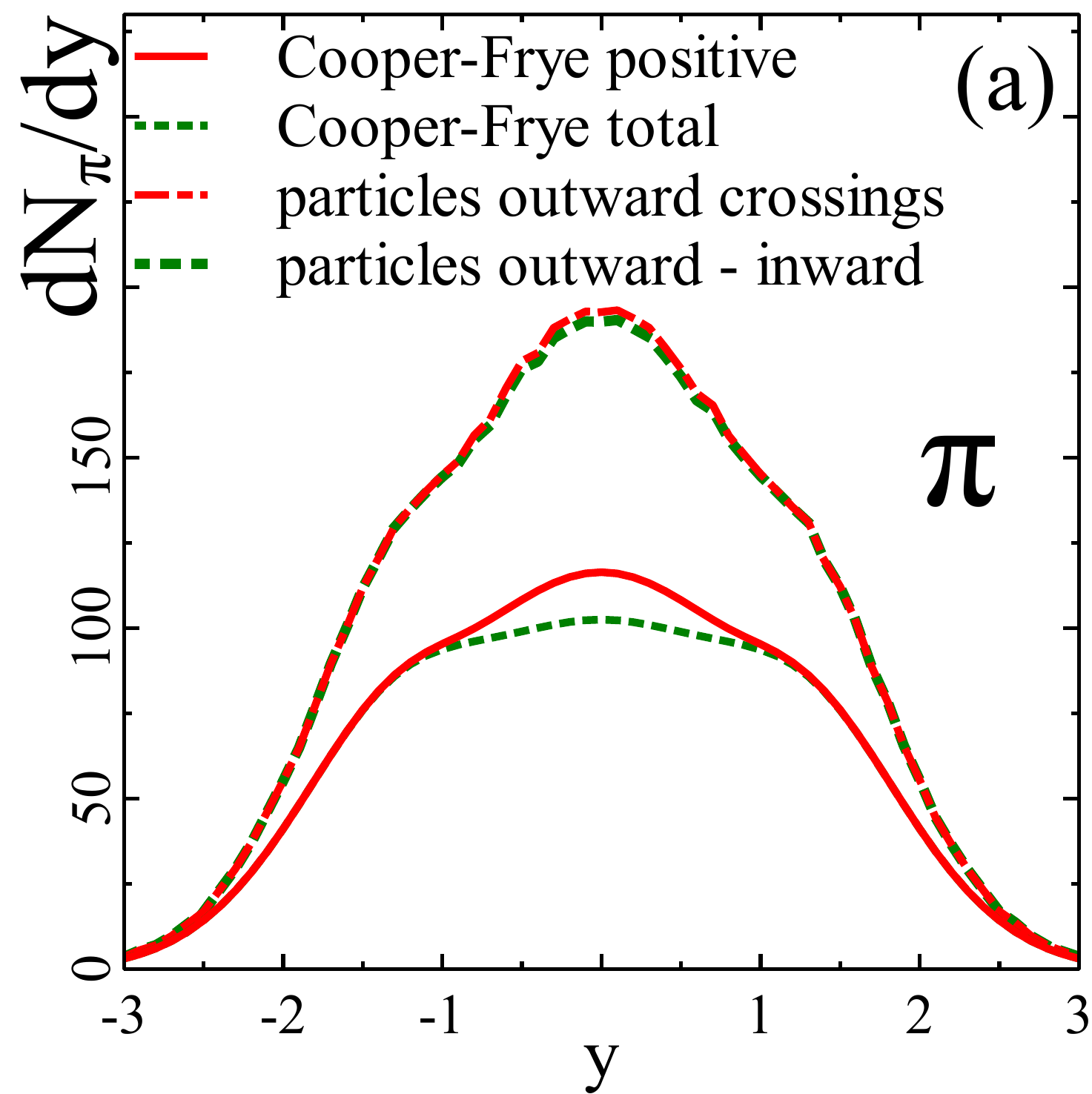}
\includegraphics[width=5cm]{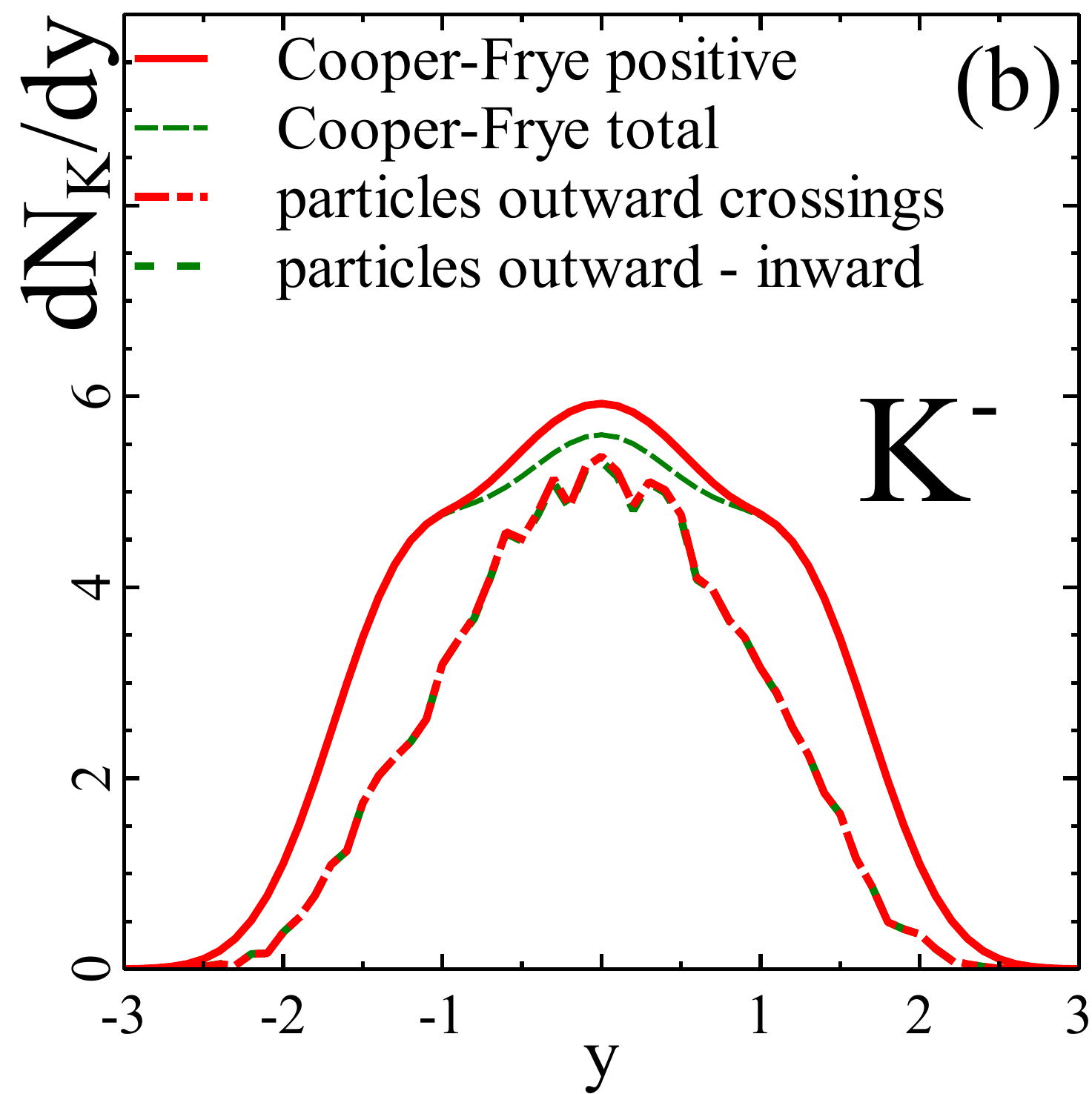}
\includegraphics[width=5cm]{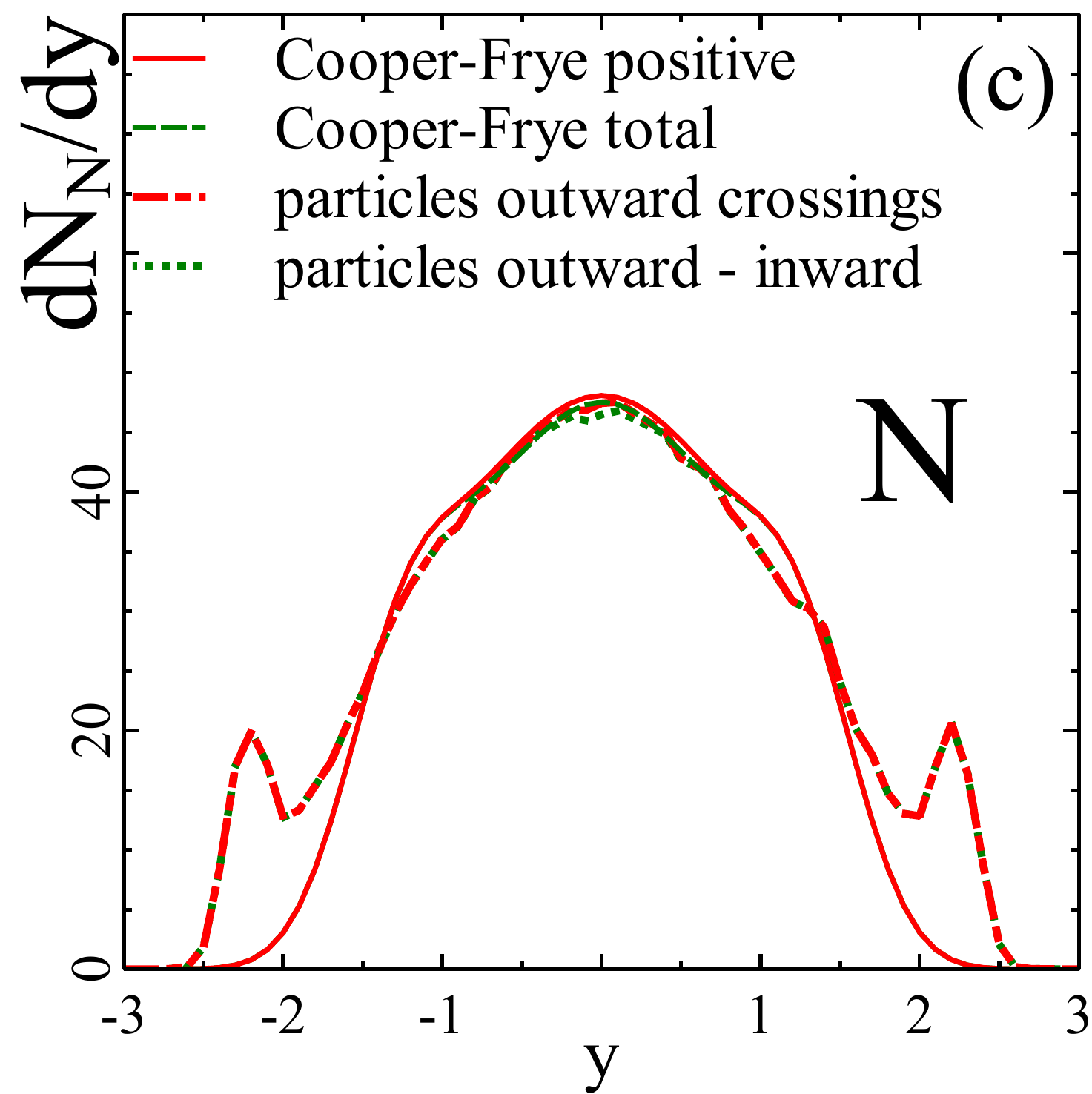}
\caption{(Color online) Rapidity distribution of identified particles
  obtained from Cooper-Frye formula on the surface $\Sigma$ and from
  explicit counting of particles that cross the same surface. Positive
  contributions and the net distribution, \emph{i.e.},
  positive-negative, are shown separately. $E_{\rm lab} = 40A$ GeV,
  central Au+Au collisions.}
\label{Fig:spectra}
\end{figure}

In Fig.~\ref{Fig:spectra} we compare rapidity spectra of identified
particles in $E_{\rm lab} = 40A$ GeV Au+Au collisions obtained by
Cooper-Frye calculation and by counting of the microscopic
particles. Even though, we are showing the results only for one
collision energy, all results are qualitatively the same at all other
energies. If UrQMD is close to equilibrium on a surface at
$\epsilon_c = 0.3$ GeV/fm$^3$, both approaches should yield similar
distributions. At midrapidity this is the case for nucleons, and with
a lesser accuracy for kaons. $\Delta$'s, $\Lambda$'s, $\rho$'s and
$\eta$'s which are not shown in the figure depict a behavior similar
to nucleons. However, the pion yields are wildly different indicating
that pions are---and thus the entire system is---far away from
chemical equilibrium at least. To cancel the effect of
non-equilibrium and to make the differences in momentum distributions
visible we consider not the absolute value of the negative
contributions, but the ratio of negative to positive ones,
$(dN^-/dy)/(dN^+/dy)$ or $(dN^-/dp_T)/(dN^+/dp_T)$. From
Fig. \ref{Fig:spectra} it is also apparent that the magnitude of the
negative contributions is always small compared to the positive ones
as expected.

\begin{figure}[htp]
\includegraphics[width=7cm]{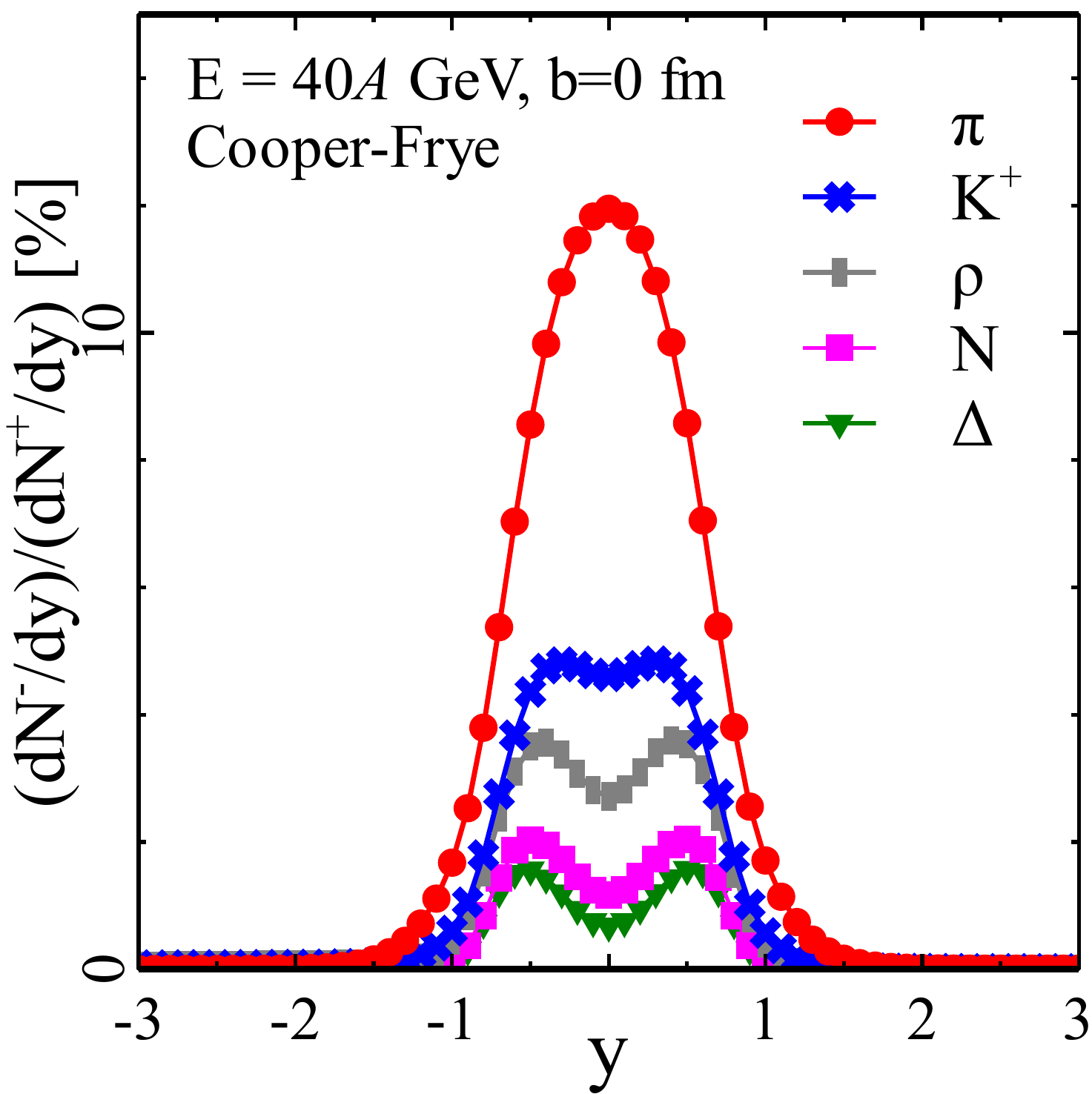}
\caption{(Color online) Rapidity distribution of the ratio of negative
  to positive contributions for different hadron species: pions
  (circles), $K^{+}$ (crosses), $\rho$ (bars), nucleons (rectangles), 
  and deltas (triangles).  Cooper-Frye calculation in central Au+Au
  collisions at $\Elab = 40A$ GeV.}
\label{Fig:neg_contr_pmass}
\end{figure}

The dependence of the ratio $(dN^-/dy)/(dN^+/dy)$ on the hadron type
is illustrated in Fig.~\ref{Fig:neg_contr_pmass} by the Cooper-Frye
results. Since for all cases, the microscopic negative contributions
of backstreaming particles are much smaller than the Cooper-Frye ones
we concentrate on showing the maximal effect. Surface temperature and
velocity profiles are identical for all hadrons, so the plot
demonstrates first of all the effect of particle mass. One can see
that the average value of $(dN^-/dy)/(dN^+/dy)$ decreases with
particle mass. This can be understood by considering a small volume of
fluid in its rest frame, and a space-like surface moving through it
with a velocity $0<v_{surf}<c$ so that lower density, \emph{i.e.},
outside, is in the negative direction. To be counted as a negative
contribution, a particle must enter the fluid, and thus have a larger
velocity than the surface. Average thermal velocity decreases with
increasing mass, and therefore the heavier the particle, the fewer of
them cross the surface inward. Since relative negative contributions
for pions are several times larger than for other hadrons we will
consider only pions in the following.

\begin{figure}[htp]
\includegraphics[width=7cm]{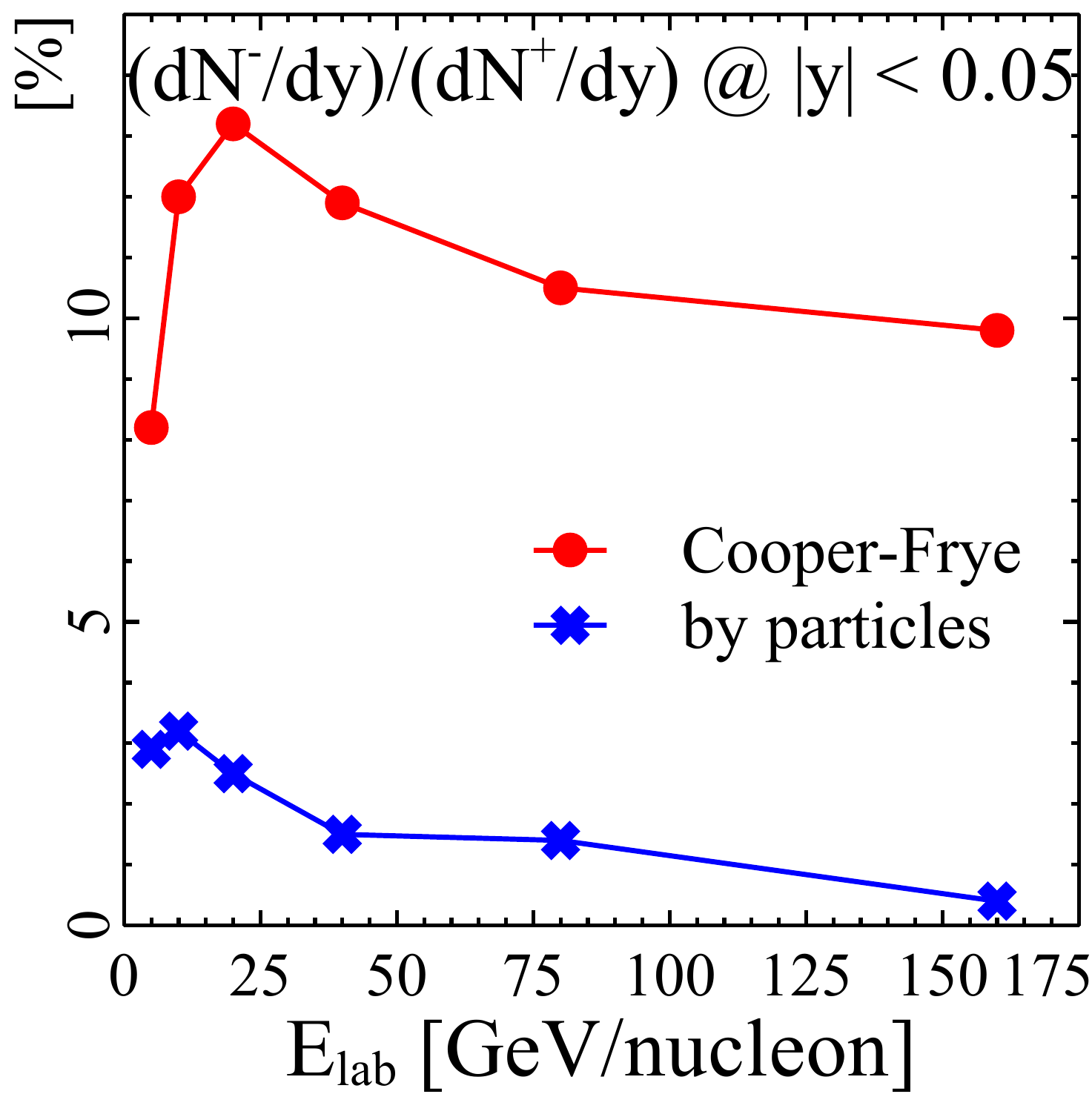}
\caption{(Color online) The ratio of negative to positive
  contributions on the $\epsilon(t,x,y,z) = \epsilon_c = 0.3$
  GeV/fm$^3$ surface for pions at midrapidity in central Au+Au
  collisions at various collision energies. Circles depict Cooper-Frye
  result and rectangles the explicit counting of UrQMD particles.}
\label{Fig:neg_contr_Ecoll}
\end{figure}

As could be seen in Fig.~\ref{Fig:spectra}, imposing equilibrium for
Cooper-Frye calculation leads to significantly larger negative to
positive contribution ratio at midrapidity than the counting of UrQMD
particles. As shown in Fig.~\ref{Fig:neg_contr_Ecoll} this holds for
all the energies we have considered, showing that the system is out of
not only chemical, but also of kinetic equilibrium. Either the
collective flow velocity of pions is different from the collective
velocity of other particles \cite{Sorge:1995pw,Pratt:1998gt} or the
dissipative corrections to pion distribution are very large. We have
also checked that the relative microscopic negative contributions are
much smaller in UrQMD at all centralities, for all particle species,
and on isosurfaces of energy density $\epsilon_c = 0.3$ and 0.6
GeV/fm$^3$.

On the other hand, the trend as a function of collision energy in
Cooper-Frye and UrQMD calculations is the same: both curves have a
maximum at 10-20$A$ GeV and then decrease with increasing energy. This
behavior is a result of a complicated interplay of several factors:
temperature, relative velocities between surface and fluid, and
relative amounts of volume and surface emission, \emph{i.e.}, emission
from the time- and space-like parts of the surface. To gain some
insight we consider all these factors separately. The same argument
used to explain the sensitivity of negative contributions to particle
mass, explains why larger temperature leads to larger negative
contributions. Temperature on the constant density surface grows with
increasing collision energy (see Fig.~\ref{Fig:surf}), which would
lead one to expect an increase of negative contributions with
increasing collision energy. On the other hand, larger relative
velocity between the fluid and surface reduces the negative
contributions (again the same argument), and we see that the average
relative velocity increases with increasing collision energy. Finally,
as argued when discussing Fig.~\ref{Fig:surf}, we have seen that the
larger the collision energy, the larger the fraction of volume
emission. Which, as mentioned, reduces the negative contributions.

\begin{figure}[htp]
\includegraphics[width=7cm]{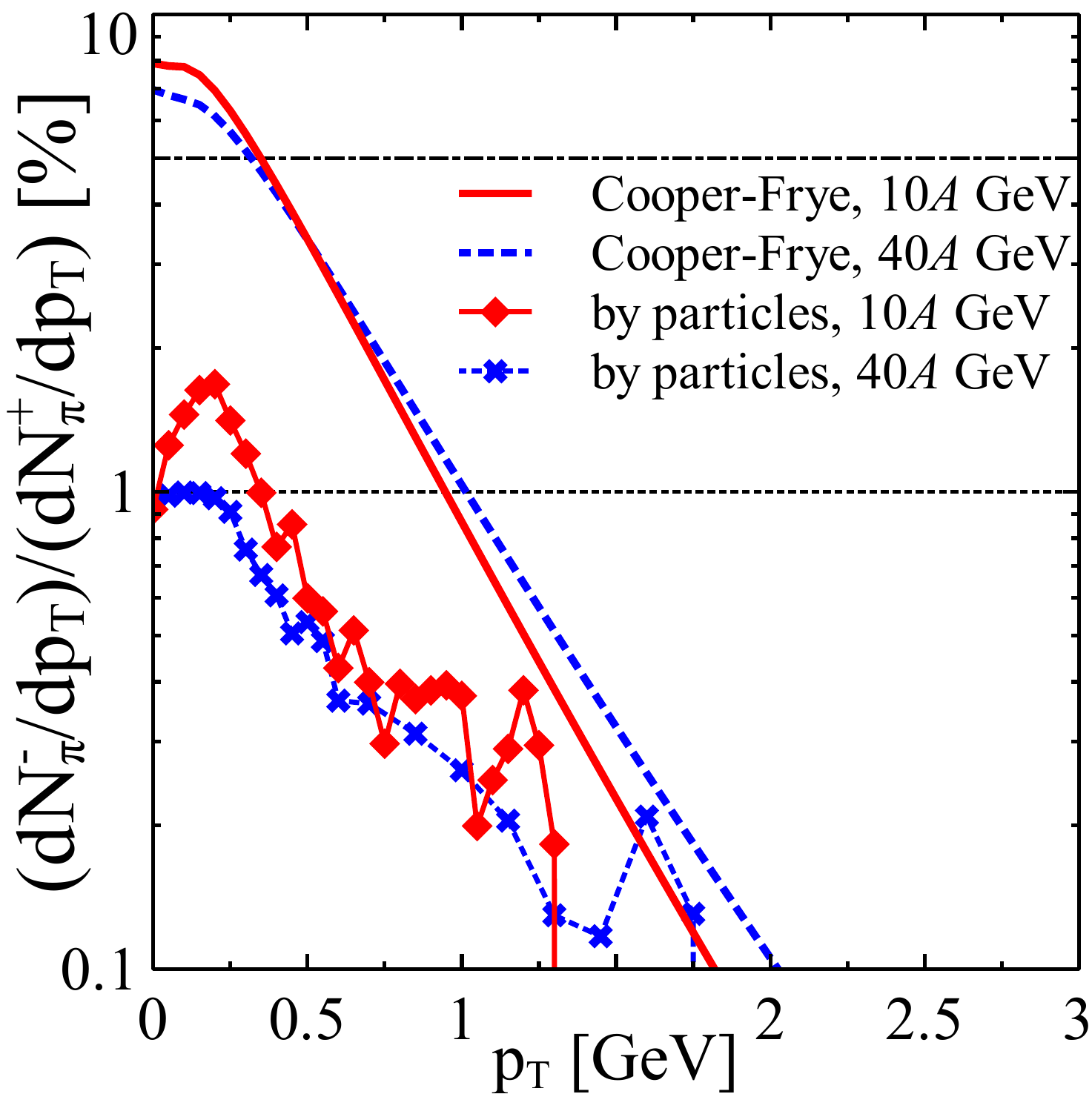}
\caption{(Color online) The ratio of negative to positive pion
  contributions as a function of transverse momentum at midrapidity in
  central Au+Au collisions at $E_{\rm lab} = 5$, 10, 20, 40, 80$A$ GeV.}
\label{Fig:neg_contr_pt}
\end{figure}

It is instructive to evaluate the negative contributions as function
of transverse momentum $p_T$ as well, as shown in
Fig.~\ref{Fig:neg_contr_pt} for Cooper-Frye calculation and "by
particles". One can see that the largest negative contributions are
located at small $p_T$, which means that one can reduce the
uncertainty caused by the negative contributions by a low $p_T$
cut. Also as a function of transverse momentum, the amount of
microscopically backward streaming particles is much smaller than in
an equilibrium scenario.

\begin{figure}[htp]
\includegraphics[width=7cm]{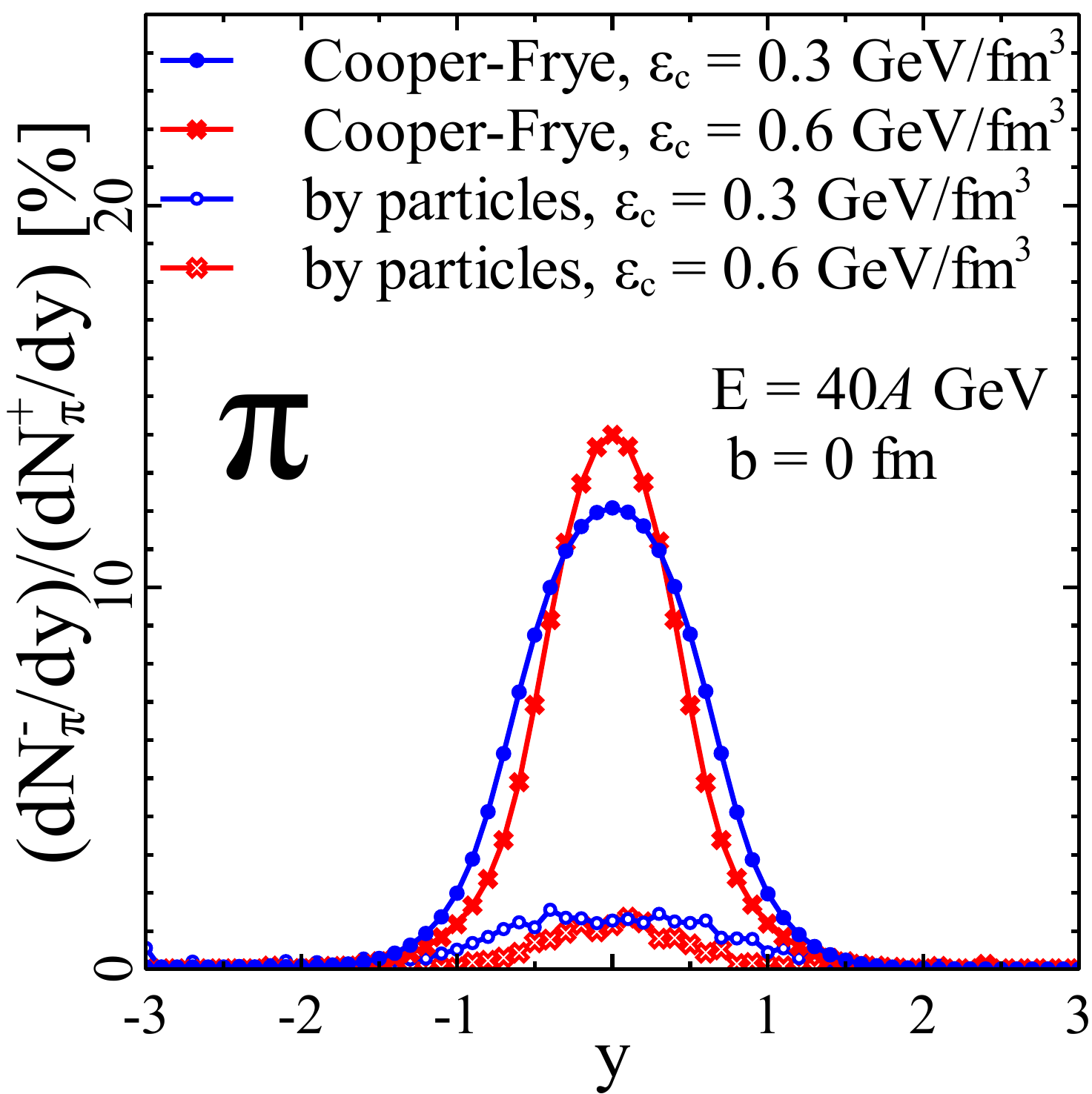}
\caption{(Color online) Rapidity distribution of the ratio of negative
  to positive contributions for pions on 
  $\epsilon(t,x,y,z) = \epsilon_c = 0.3$ GeV/fm$^3$ (circles) and
  $\epsilon_c = 0.6$ GeV/fm$^3$ (crosses) surfaces in central Au+Au
  collisions at $\Elab = 40\ A$ GeV. Full symbols correspond to
  Cooper-Frye calculation and open symbols to explicit counting of
  UrQMD particles.}
\label{Fig:neg_contr_e0}
\end{figure}

When discussing Fig.~\ref{Fig:neg_contr_Ecoll} we mentioned that,
independent of the energy density of the surface, the negative
contributions are much smaller when counting the UrQMD
particles. Furthermore, in Cooper-Frye calculations the strength of
the negative contributions depends on the value of $\epsilon_c$ where
the distributions are evaluated as shown in
Fig.~\ref{Fig:neg_contr_e0}. Larger $\epsilon_c$ leads to larger
negative contribution at midrapidity and lower at back- and forward
rapidities. This result arises from interplay of two factors: larger
temperature and smaller average $v_{rel}$ for larger energy
density. Quite surprisingly the negative contributions evaluated by
counting the UrQMD particles is almost independent of the value of
$\epsilon_c$. This indicates that even in much higher temperature
$T\sim 155$--160 MeV the microscopic system is not fully thermalised.

\begin{figure}[htp]
\includegraphics[height=7cm]{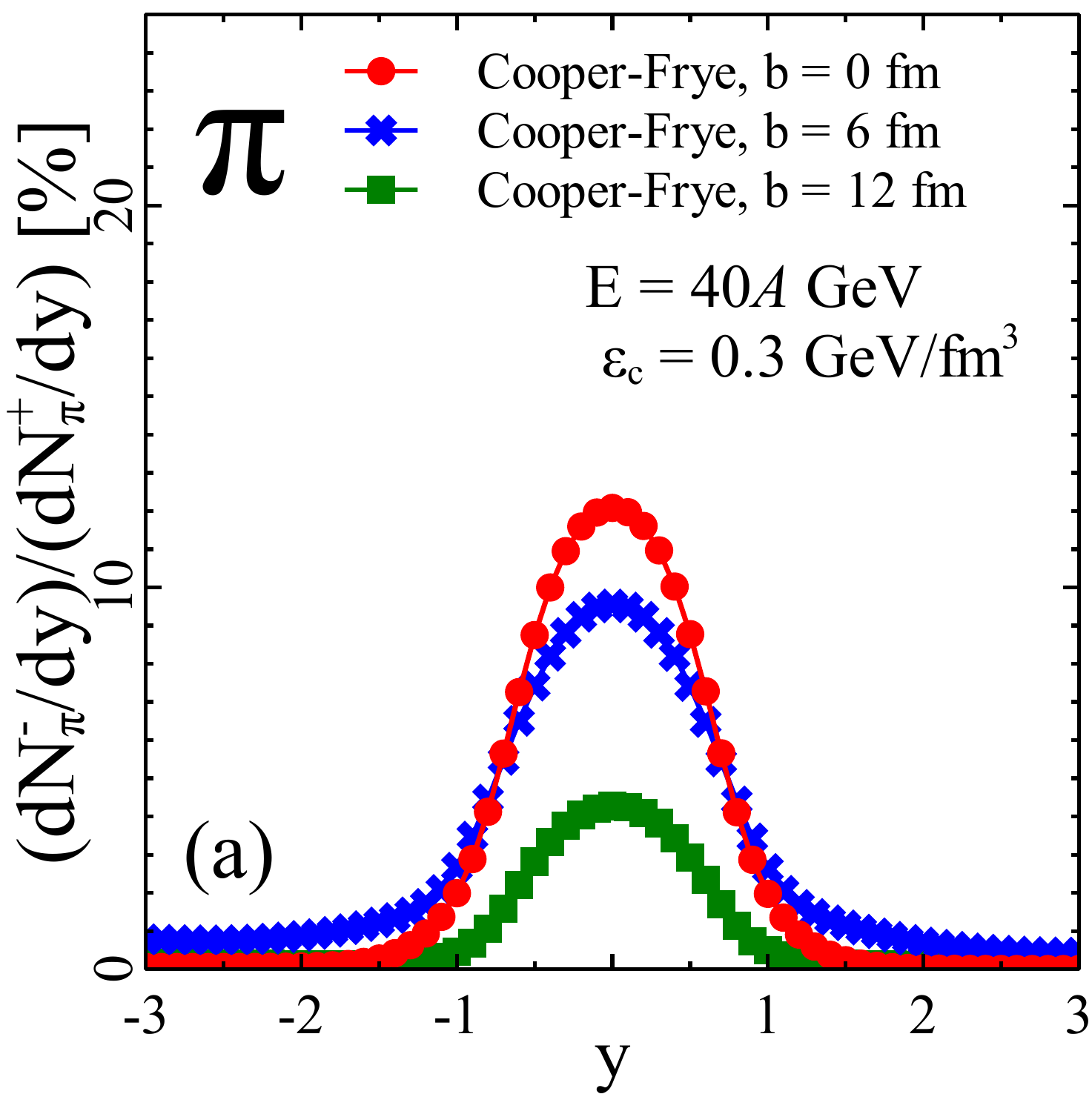}
\includegraphics[height=7cm]{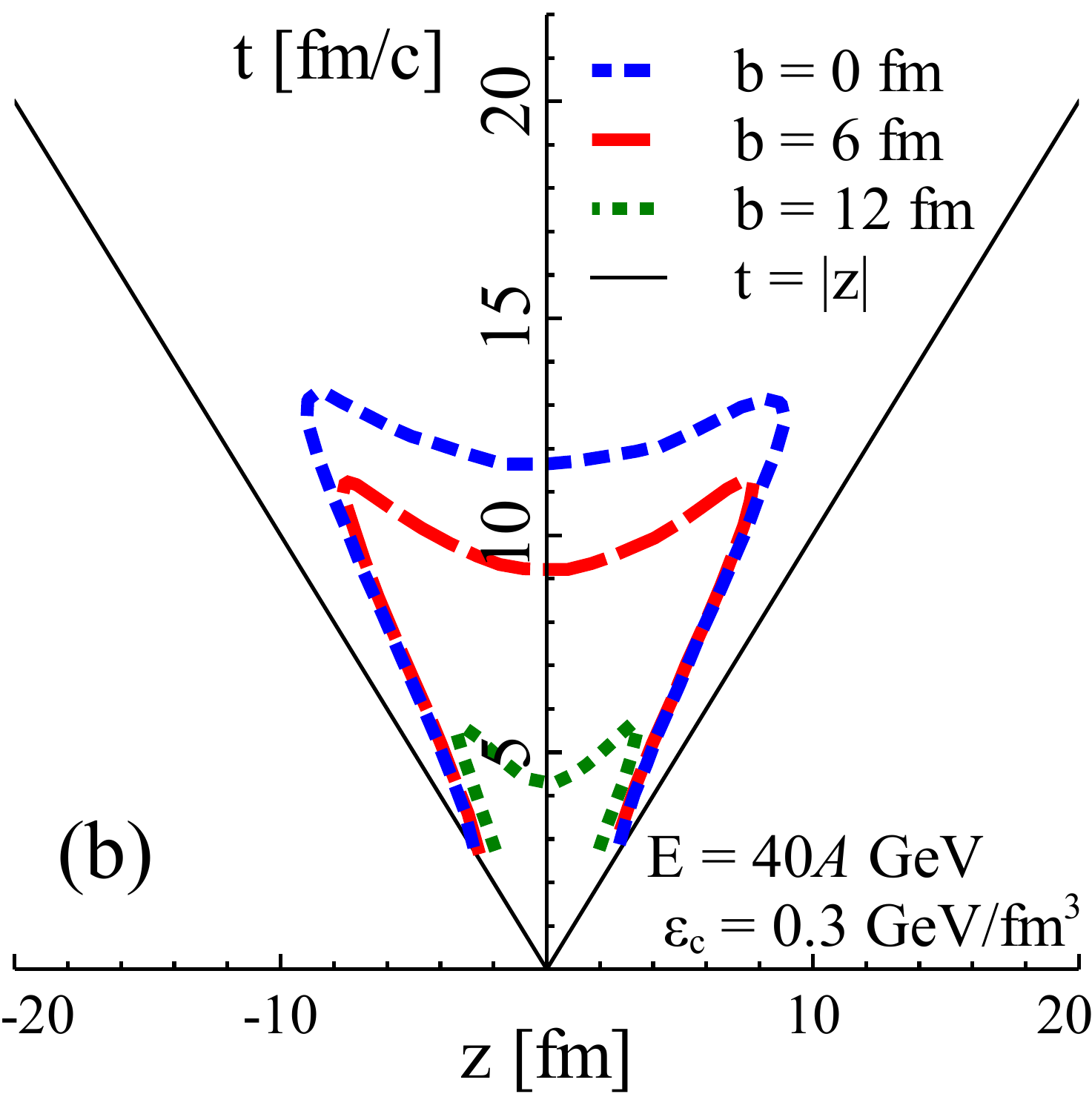}
\caption{(Color online) Upper panel: Rapidity distribution of the
  ratio of negative to positive contributions for pions in Au+Au
  collisions at $\Elab = 40\ A$ GeV at various centralities: $b=0$
  (circles), $b = 6$ fm (crosses) and $b = 12$ fm (rectangles). Lower
  panel: hypersurfaces along the z axis in the same collisions at the
  same centralities.}
\label{Fig:neg_contr_b}
\end{figure}

Dependence of the contribution ratio on centrality is shown in
Fig.~\ref{Fig:neg_contr_b}. The negative contributions decrease with
decreasing centrality because the more peripheral the collision, the
larger the fraction of time-like hypersurface elements. This behavior
is illustrated in the right panel of Fig.~\ref{Fig:neg_contr_b}. In
the limit of very peripheral collisions the lifetime of the system
becomes zero, and thus the surface is time-like everywhere and there
are no negative contributions at all. Temperature and relative
velocities appear to be less important factors in this case than
relative amount of time-like and space-like hypersurface elements.

Let us finally compare our results to previous studies. In
\cite{ref:PetHuo12} negative contributions were evaluated on the
$\epsilon = 0.3$ GeV/fm$^3$ transition surface of a hybrid model at
SPS and RHIC energies---$E_{\rm lab} = 160A$ GeV and
$\sqrt{s_\mathrm{NN}} = 200$ GeV, respectively---and found to be
around $(dN^{-}_{\pi}/dy)/(dN^{+}_{\pi}/dy) \simeq$ 13\% and 9\% at
$y=0$. The negative contributions for 160$A$ GeV are slightly larger
than in our calculation. The reason for this discrepancy lies in the
difference of the velocity profiles on the hypersurfaces: In
hydrodynamics the average relative velocity between flow and surface
is smaller than in our transport-based approach, which leads to larger
negative contributions.

\section{Conclusions}

We have investigated negative Cooper-Frye contributions and
backscattering using a coarse-grained molecular dynamics
approach. Au+Au collisions at $\Elab = 5$--$160\ A$ GeV energies have
been simulated using UrQMD, and a hypersurface $\Sigma$ of constant
Landau rest frame energy density has been constructed. On this surface
we have calculated two quantities: The ratio of Cooper-Frye negative
to positive contributions, which assumes local thermal equilibrium,
and the ratio of UrQMD particles crossing $\Sigma$ inward to crossing
$\Sigma$ outward, which assumes no equilibrium.

We found that at all collision energies the ratio of inward to outward
moving particles calculated counting the UrQMD particles is much
smaller than the same ratio calculated assuming equilibrium,
\emph{i.e.}, the Cooper-Frye negative to positive ratio. This finding
poses a question to the construction of hybrid models, and the
treatment of freeze-out in hydrodynamical models: If the cascade leads
to distributions nowhere near equilibrium, how are the hydrodynamical
and cascade stages to be connected in a consistent fashion? On the
other hand, this result shows that an ideal fluid dynamics hybrid
approach contains the worst case scenario for negative contributions
and even then they are on the order of max.~$15\%$ for the pion yield
at midrapidity. What remains to be seen, however, is whether we could
get closer to the UrQMD result if we allowed dissipative corrections
to the distribution function of Cooper-Frye, or whether the deviations
from equilibrium are so large that dissipative expansion is not
feasible.

The largest observed impact of negative contributions is to pion
rapidity spectrum at midrapidity in central collisions. In thermally
equilibrated Cooper-Frye calculations it constitutes 8--13\%, but only
0.5--4\% in the counting of UrQMD particles. The Cooper-Frye value
roughly agrees with the values obtained previously for hydrodynamics
at 160 GeV. We found several systematic features in these ratios. They
are smaller for larger hadron mass and therefore largest for
pions. The relative negative contributions decrease as a function of
collision energy and by going from central to peripheral
collisions. On the other hand, they increase if a higher energy
density is chosen as a surface criterion. The small scale structures
on the surface, its ``lumpiness'', play a significant role: If the
surface is not smooth enough both ratios can increase
dramatically. Therefore, an interesting future study could be to
compare single fluctuating events to the averaged result.

\begin{acknowledgments}
  This work was supported by the Helmholtz International Center for
  the Facility for Antiproton and Ion Research (HIC for FAIR) within
  the framework of the Landes-Offensive zur Entwicklung
  Wissenschaftlich-Oekonomischer Exzellenz (LOEWE) program launched by
  the State of Hesse. DO and HP acknowledge funding of a Helmholtz
  Young Investigator Group VH-NG-822 from the Helmholtz Association
  and GSI, and PH by BMBF under contract no.~06FY9092. Computational
  resources have been provided by the Center for Scientific Computing
  (CSC) at the Goethe University of Frankfurt.
\end{acknowledgments}

\end{document}